\documentclass[11pt]{cernrep}

\usepackage{url}
\usepackage{graphicx}
\usepackage{amsmath}
\usepackage{amssymb}

\usepackage{makeidx}

\allowdisplaybreaks[2]

\def\d{\mathrm{d}}

\begin{document}

\title{Constraining the $\rho$ meson wavefunction\footnote{~~Proceedings
Contribution for the workshop \textit{Gluons and the quark sea at high
energies: distributions, polarization, tomography}, Institute for Nuclear
Theory, Seattle, U.S. September 13 to November 19, 2010.}}
\author{J. R. Forshaw\footnote{~~jeff.forshaw@manchester.ac.uk}~~$\&$  R.
Sandapen\footnote{~~ruben.sandapen@umoncton.ca}}
\institute{University of Manchester $\&$ Universit\'e de
Moncton\\
MAN/HEP/2010/24
}

\maketitle





%


\vspace{2\baselineskip}


\centerline{\bf Abstract}
\vspace{0.7\baselineskip}
\parbox{0.9\textwidth}{Diffractive $\rho$ meson production has been identified
as one of the important
processes where saturation can be probed at a future Electron-Ion Collider
(EIC). A source of uncertainty in making predictions for this process lies
within the assumed form of the meson light-cone wavefunction. We report
here the results of reference~\cite{Forshaw:2010py} where a Regge-inspired
dipole
model was used to extract this wavefunction as well as the corresponding leading
twist-$2$ Distribution Amplitude from the current accurate HERA data. In
addition, we shall check the robustness of the conclusions in
reference~\cite{Forshaw:2010py} by using alternative Colour Glass Condensate
dipole
models. }


\section{Introduction}

In the dipole model~\cite{Nikolaev:1990ja,Mueller:1994jq}, the 
imaginary part of the amplitude  for diffractive $\rho$ production is written 
as~\cite{Marquet:2007qa}
\begin{equation}
\Im \mbox{m} \mathcal{A}_{\lambda} (s,t;Q^2) =
\sum_{h,\bar{h}} \int \d^2  \mathbf{r} \d z
\Psi^{\gamma^*,\lambda}_{h,\bar{h}} (r,z;Q^2)
\Psi^{\rho,\lambda}_{h,\bar{h}}(r,z)^{*}
e^{-iz\mathbf{r}.\mathbf{\Delta}} \mathcal{N}(x,\mathbf{r},\mathbf{\Delta}) 
\label{non-forward-amplitude}
\end{equation}
where $t=-|\mathbf{\Delta}|^2$. In a standard
notation~\cite{Forshaw:2010py,Marquet:2007qa,Forshaw:2003ki},
$\Psi^{\gamma^*,\lambda}_{h,\bar{h}}$ and $\Psi^{\rho,\lambda}_{h,\bar{h}}$ are
the light-cone wavefunctions of the photon and the $\rho$ meson respectively
while $\mathcal{N}(x,\mathbf{r},\mathbf{\Delta})$ is the imaginary part of the
dipole-proton elastic scattering amplitude. The energy dependence of the latter
is via the dimensionless variable $x$ taken here to
be
\begin{equation}
x = (Q^2 + 4m_f^2)/(Q^2+ s)  
\end{equation}
where $m_f$ is a phenomenological light quark mass.\footnote{Here we shall take
$m_f=0.14~\mbox{GeV}$ i.e the 
value used when extracting the dipole
cross-section from $F_2$ data.} Setting $t=0$ in equation
\eqref{non-forward-amplitude}, we obtain the forward
amplitude used in reference~\cite{Forshaw:2010py}, i.e
\begin{equation}
\left.\Im \mbox{m} \, \mathcal{A}_\lambda(s,t;Q^2)\right|_{t=0} = s \sum_{h,
\bar{h}}
\int \d^2 {\mathbf r} \; \d z \; \Psi^{\gamma,\lambda}_{h, \bar{h}}(r,z;Q^2)
\hat{\sigma}(x,r) \Psi^{\rho,\lambda}_{h, \bar{h}}(r,z)^*
\label{forward-amplitude}
\end{equation}
where we have used the optical theorem to introduce the dipole cross-section
\begin{equation}
\hat{\sigma}(x,r)=\frac{\mathcal{N}(x,r,\mathbf{0})}{s}\;.
\label{dipole-xsec-optical}
\end{equation}
Note that since the momentum transfer $\mathbf{\Delta}$ is Fourier conjugate to
the impact parameter $\mathbf{b}$, the dipole cross-section at a given energy
is simply the $b$-integrated dipole-proton scattering amplitude, i.e 
\begin{equation}
\hat{\sigma}(x,r)=\frac{1}{s}\int
\d^2 \mathbf{b} \mathcal{N}(x,r,\mathbf{b}) \;.
\label{dipole-xsec-b-int}
\end{equation}
This dipole
cross-section can be extracted from
the $F_2$ data since
\begin{equation}
F_2(x,Q^2)  \propto \int \d^2 \mathbf{r} \d z |\Psi_{\gamma^*}(r,z;Q^2)|^2
\hat{\sigma}(x,r)
\end{equation}
and the photon's light-cone wavefunctions are known in QED, at least for large
$Q^2$. The
$F_2$-constrained dipole cross-section can then be used
to predict the imaginary part of the forward amplitude 
for diffractive $\rho$ production and thus the forward differential
cross-section, 
\begin{equation}
\left. \frac{d\sigma_{\lambda}}{dt} \right|_{t=0} =\frac
{1}{16\pi} (\Im\mathrm{m} \mathcal{A}_\lambda(s,0))^2 \; (1 + \beta_\lambda^2)~,
\label{gammap-xsec}
\end{equation}
where $\beta_\lambda$ is the ratio of real to imaginary parts of the
amplitude and is computed as in reference~\cite{Forshaw:2010py}. The
$t$-dependence
can be assumed to be the
exponential dependence as suggested by experiment~\cite{Chekanov:2007zr}:
\begin{equation}
\frac{d\sigma_{\lambda}}{dt}= \left. \frac{d\sigma_{\lambda}}{dt} \right|_{t=0}
\times
\exp(-B|t|)
\label{exponential-t}
\end{equation}
where
\begin{equation}
B=N\left(
  14.0 \left(\frac{1~\mathrm{GeV}^2}{Q^2 + M_{\rho}^2}\right)^{0.2}+1\right)
\label{Bslope}
\end{equation}
with $N=0.55$ GeV$^{-2}$. After integrating over $t$, we can compute the total
cross-section $\sigma=\sigma_L + \epsilon \sigma_T$ 
which is measured at HERA.\footnote{To compare with the HERA data, we take
$\epsilon=0.98$.} 

Presently several dipole
models~\cite{Forshaw:2004vv,Watt:2007nr,Soyez:2007kg,Bartels:2002cj,
Kowalski:2006hc} 
are able to fit the current HERA $F_2$ data and 
there is evidence that the data prefer those incorporating
some form of saturation~\cite{Motyka:2008jk}. We can use the $F_2$-constrained
dipole cross-section in order to
extract the $\rho$ light-cone wavefunction using the current precise HERA data
~\cite{Chekanov:2007zr,Aaron:2009xp}. This
has recently been done in reference~\cite{Forshaw:2010py} using the
Regge-inspired  FSSat dipole
model~\cite{Forshaw:2004vv} and we shall report the results of
this work here. In addition, we repeat the analysis 
using two alternative models~\cite{Soyez:2007kg,Watt:2007nr,Kowalski:2006hc}
both
based on the original Colour Glass Condensate (CGC) model~\cite{Iancu:2003ge}.
They differ from the original CGC model by including the
contribution of charm quarks when fitting to the $F_2$
data. Furthermore in one of them~\cite{Soyez:2007kg,Watt:2007nr}, the anomalous
dimension  $\gamma_s$ is treated as an additional free parameter instead of
being fixed to its LO BFKL value of $0.63$. We shall refer to
these models as CGC[$0.74$] and CGC[$0.63$] models  where the number in the
square brackets stands for the fitted and fixed value of the anomalous
dimension respectively. For both models, we use the set of fitted parameters
given in reference~\cite{Watt:2007nr}. All
three models, i.e FSSat, CGC[$0.63$] and CGC[$0.74$] account for saturation
although in a $b$- (or equivalently $t$-) independent way. Indeed, at
a given energy, the dipole cross-section is equal to the forward
dipole-proton amplitude given by equation \eqref{dipole-xsec-optical} or to the
$b$-integrated
dipole proton amplitude given by equation \eqref{dipole-xsec-b-int}. Finally 
we note that all three dipole models we consider here give a good description of
the diffractive structure function, i.e $F_2^{D(3)}$
data~\cite{Marquet:2007nf,Forshaw:2006np}.

\section{Fitting the HERA data}

Previous work~\cite{Forshaw:2003ki,Marquet:2007qa,Watt:2007nr}
has shown that a reasonable assumption for the
scalar part of the light-cone wavefunction for the $\rho$ is of the form
\begin{eqnarray}
\phi^{{\mathrm{BG}}}_\lambda(r,z) &=&
\mathcal{N}_\lambda \;  4[z(1-z)]^{b_{\lambda}} \sqrt{2\pi R_{\lambda}^{2}} \;
\exp \left(\frac{m_f^{2}R_{\lambda}^{2}}{2}\right)
\exp \left(-\frac{m_f^{2}R_{\lambda}^{2}}{8[z(1-z)]^{b_{\lambda}}}\right) \\
\nonumber
& &\times \exp \left(-\frac{2[z(1-z)]^{b_\lambda}
r^{2}}{R_{\lambda}^{2}}\right) 
\label{boosted-gaussian} 
\end{eqnarray}
and is referred to as the 'Boosted Gaussian' (BG). This wavefunction is a
simplified version of that proposed originally
by Nemchik, Nikolaev, Predazzi and Zakharov~\cite{Nemchik:1996cw}. In the
original BG wavefunction, $b_\lambda=1$ while the parameters $R_{\lambda}$ and
$\mathcal{N}_{\lambda}$ are fixed by the leptonic decay width constraint and
the wavefunction normalization conditions~\cite{Forshaw:2010py}.

However, when the BG wavefunction is used in conjunction with either the FSSat
model or any of the CGC models, none of them is able to give a good quantitative
agreement with the current HERA $\rho$-production data.
This is illustrated by the large $\chi^2$ values in table \ref{tab:BG-chisqs}.
As shown in table \ref{tab:BG-Fits-chisqs}, this
situation is considerably improved by fitting $R_\lambda$ and $b_\lambda$ to
the leptonic decay width and HERA data.\footnote{We fit to the same data set and
with the same cuts as in reference~\cite{Forshaw:2010py}.}

\begin{table}[h]
\begin{center}
\textbf{Boosted Gaussian predictions}
\[
\begin{array} 
[c]{|c|c|}\hline
\mbox{Dipole model} & \chi^2/\mbox{data point}\\ \hline
\mbox{FSSat}& 310/75 \\ \hline
\mbox{CGC}[0.74]& 262/75 \\ \hline
\mbox{CGC}[0.63]&  401/75 \\ \hline
\end{array}
\]
\end{center}
\caption {Predictions of the $\chi^2/\mbox{data point}$ using the BG
wavefunction.}
\label{tab:BG-chisqs}
\end{table}

\begin{table}[h]
\begin{center}
\textbf{BG fits}
\[
\begin{array} 
[c]{|c|c|}\hline
\mbox{Model} & \chi^2/\mbox{d.o.f}  \\ \hline
\mbox{FSSat~\cite{Forshaw:2010py}}&  82/72 \\ \hline
\mbox{CGC}[0.74]& 64/72 \\ \hline
\mbox{CGC}[0.63] & 83/72 \\ \hline
\end{array}
\]
\end{center}
\caption {$\chi^2/\mbox{d.o.f}$ obtained when fitting $R_{\lambda}$ and
$b_{\lambda}$ to the leptonic decay width and HERA data.}
\label{tab:BG-Fits-chisqs}
\end{table}

For the FSSat and CGC[$0.63$] models, we can further improve the quality of fit
by allowing for
additional end-point enhancement in the transverse wavefunction, i.e. using a
scalar
wavefunction of the form
\begin{equation}
\phi_T (r,z)= \phi^{{\mathrm{BG}}}_T (r,z) \times [1+ c_{T}
\xi^2 + d_{T} \xi^4]
\label{EG} 
\end{equation}
where $\xi=2z-1$. The results are shown in table
\ref{tab:Enhancement-fits-chisqs}.

\begin{table}[h]
\begin{center}
\textbf{Improved fits}
\[
\begin{array} 
[c]{|c|c|}\hline
\mbox{Model} & \chi^2/\mbox{d.o.f} \\ \hline
\mbox{FSSat~\cite{Forshaw:2010py}}&  68/70 \\ \hline
\mbox{CGC}[0.63] & 67/70  \\ \hline
\end{array}
\]
\end{center}
\caption {$\chi^2/\mbox{d.o.f}$ obtained when fitting $b_{\lambda}$,
$R_{\lambda}$
$c_{T}$, $d_{T}$ the leptonic decay width and HERA data.}
\label{tab:Enhancement-fits-chisqs}
\end{table}

\begin{table}[h]
\begin{center}
\textbf{Best fit parameters}
\[
\begin{array} 
[c]{|c|c|c|c|c|c|c|}\hline
                 & R_L^2 & R_T^2 & b_L & b_T & c_T & d_T \\ \hline
\mbox{FSSat~\cite{Forshaw:2010py}}     &26.76  &27.52 &0.5665 &0.7468 &0.3317
&1.310  \\ \hline
\mbox{CGC}[0.63] &27.31  &31.92 &0.5522 &0.7289 &1.6927 &2.1457  \\ \hline
\mbox{CGC}[0.74] &26.67  &21.30 &0.5697 &0.7929  &0&0 \\ \hline
\end{array}
\]
\end{center}
\caption {Best fit parameters for each dipole model.}
\label{tab:Best-fit-params}
\end{table}

\begin{figure}
\begin{center}
\includegraphics[width=0.8\textwidth]{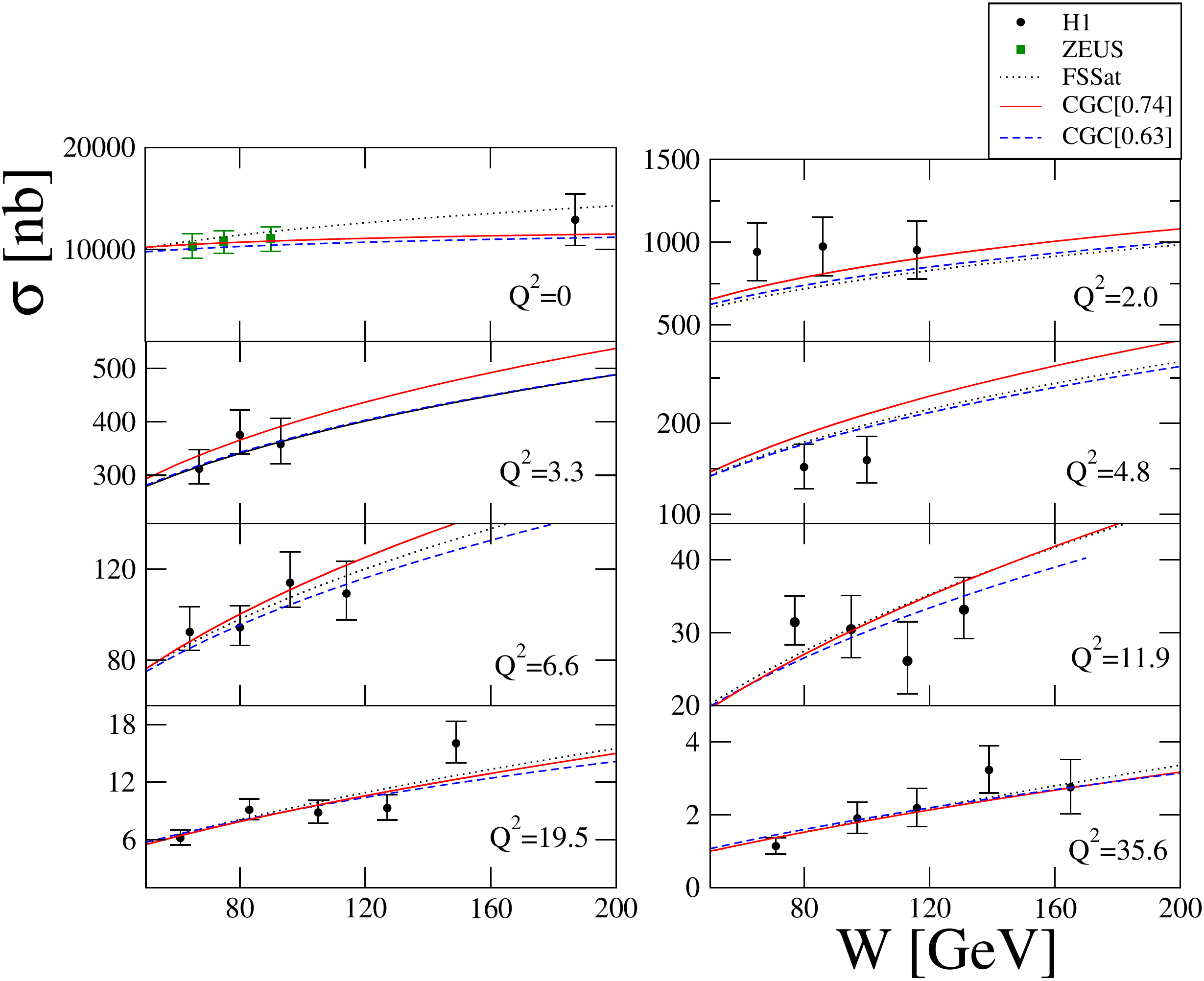}
\end{center}
\caption{\label{fig:H1-xsec} Best fits to the HERA total cross-section data.
CGC[$0.74$]: solid;
FSSat: dotted; CGC[$0.63$]: dashed.}
\end{figure}

\begin{figure}
\begin{center}
\includegraphics[width=0.8\textwidth]{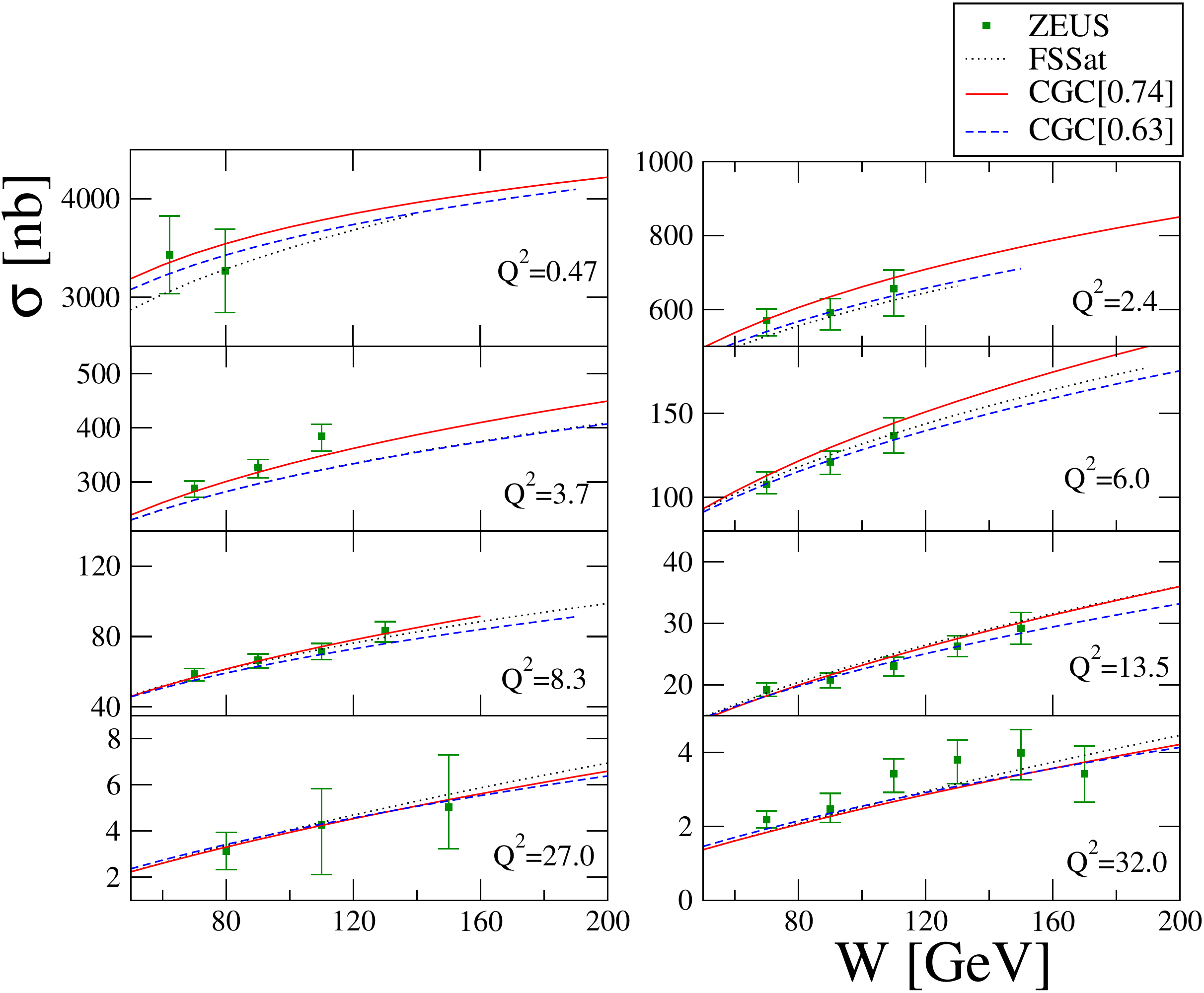}
\end{center}
\caption{\label{fig:ZEUS-xsec} Best fits to the ZEUS total cross-section data.
CGC[$0.74$]: solid;
FSSat: dotted; CGC[$0.63$]: dashed.}
\end{figure}

\begin{figure}
\begin{center}
\includegraphics[width=0.4\textwidth]{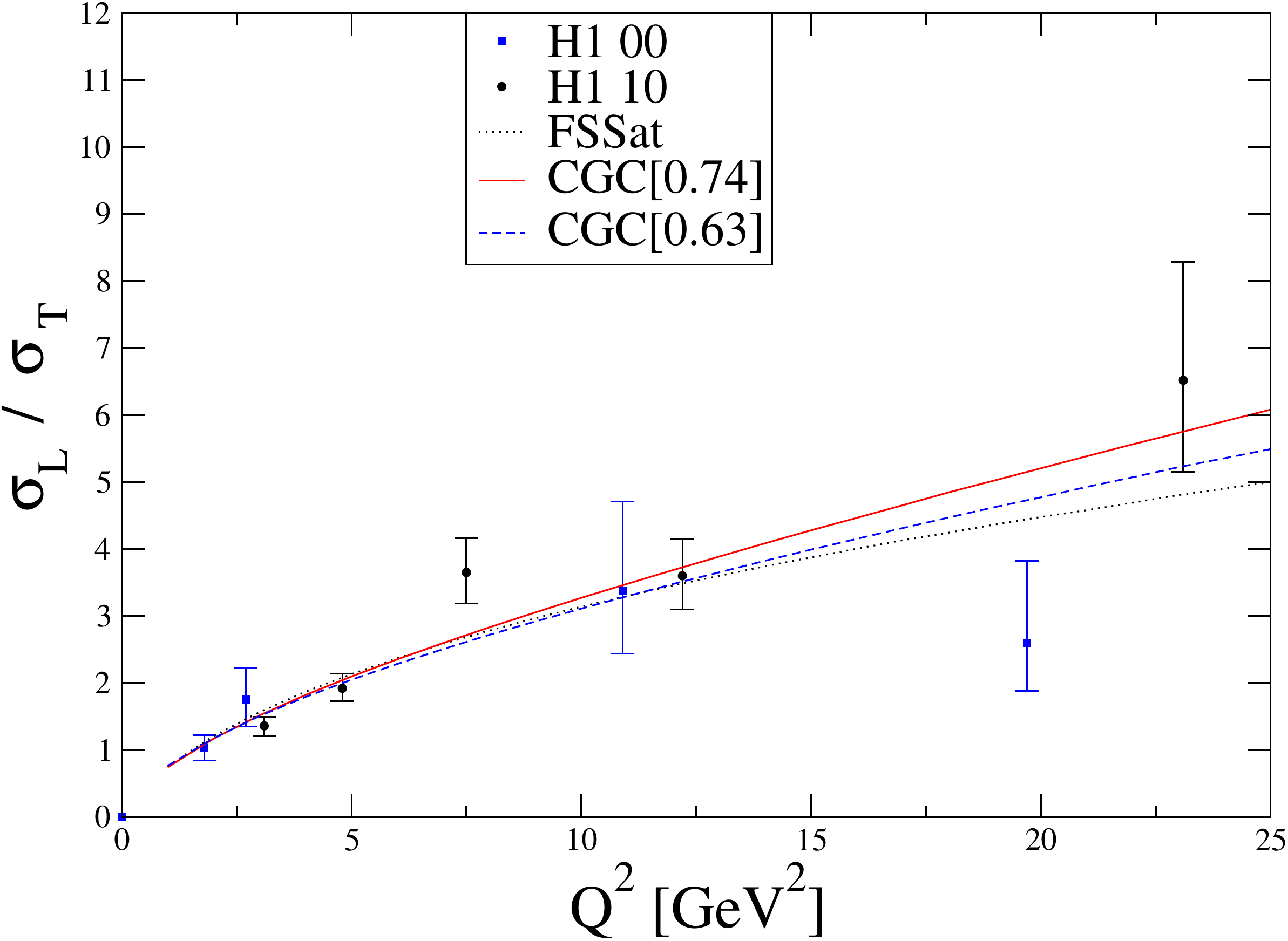}
\includegraphics[width=0.4\textwidth]{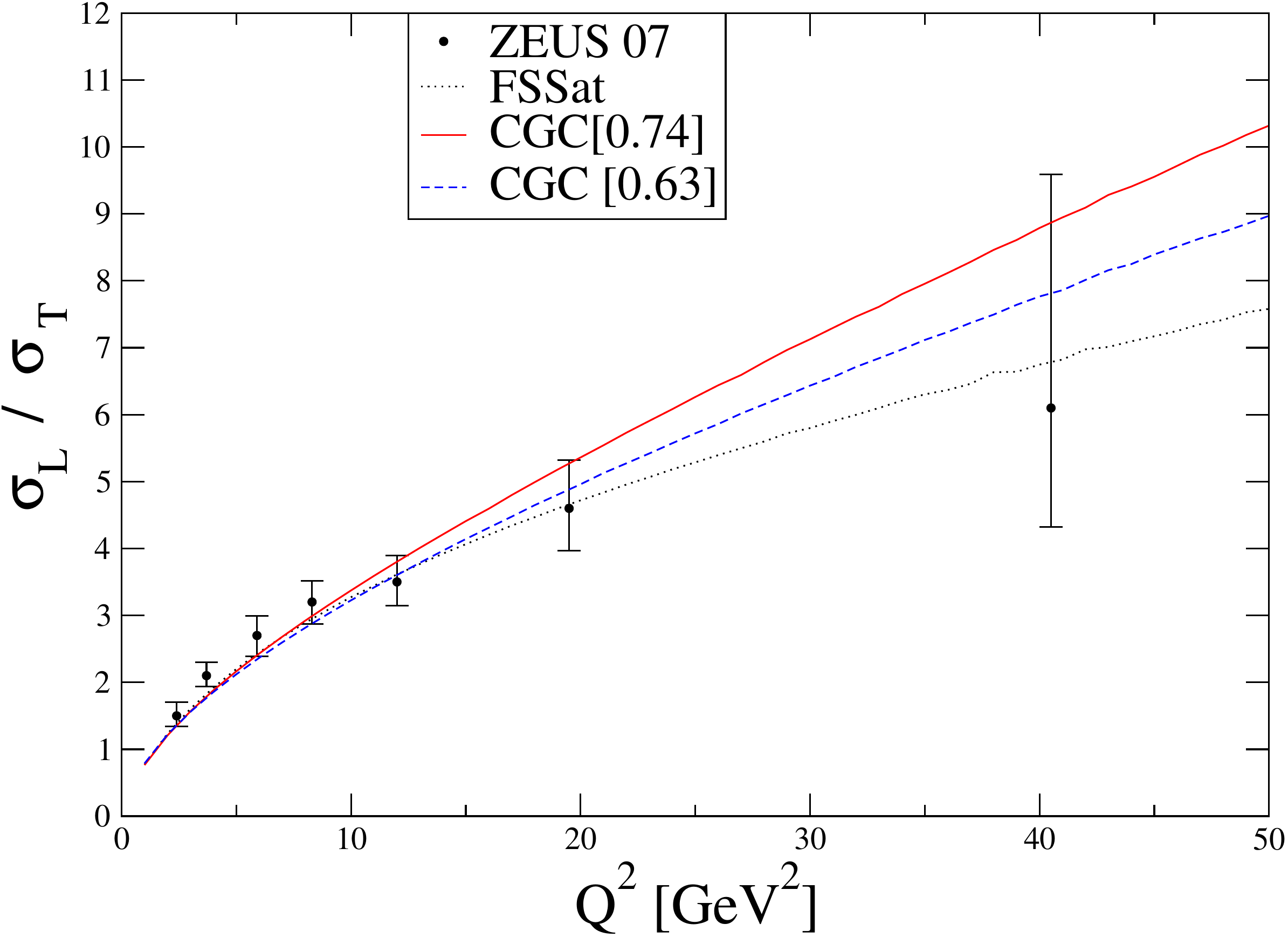}
\end{center}
\caption{\label{fig:HERA-ratio} Best fits to the $\sigma_L/\sigma_T$ data. The
H1 data are at $W=75$~GeV while the ZEUS data are at $W=90$~GeV.
CGC[$0.74$]: solid;
FSSat: dotted; CGC[$0.63$]: dashed.}
\end{figure}

\begin{figure}
\begin{center}
\includegraphics[width=0.4\textwidth]{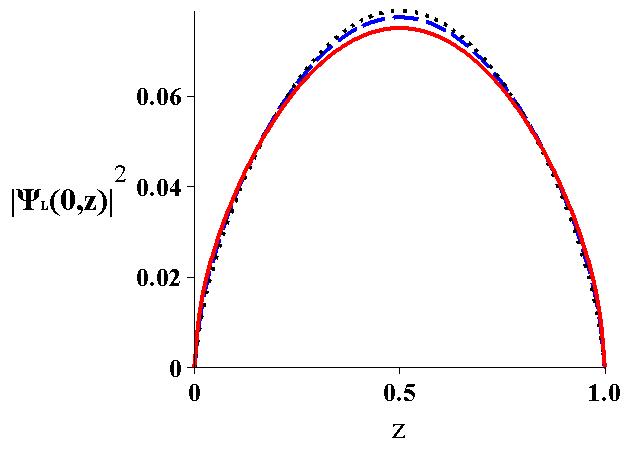}
\includegraphics[width=0.4\textwidth]{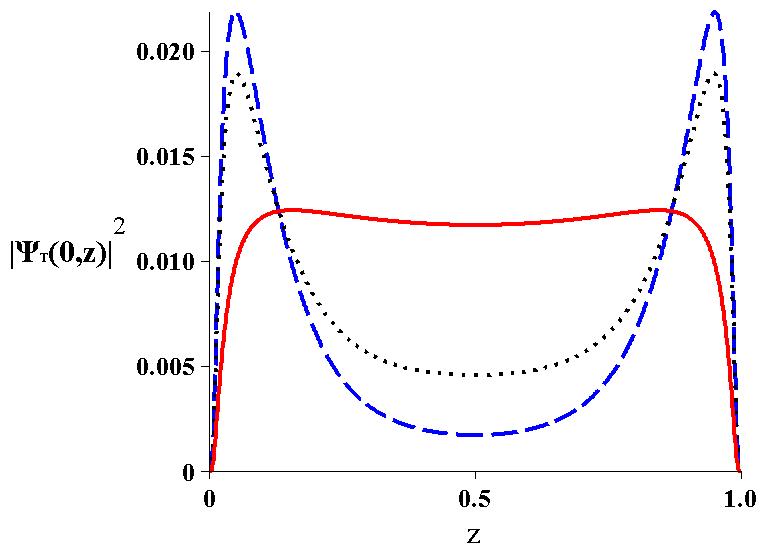}
\end{center}
\caption{\label{fig:Psisq_r0}The longitudinal (left) and transverse (right)
light-cone
wavefunctions squared at $r=0$. CGC[$0.74$]: solid; FSSat: dotted; CGC[$0.63$]:
dashed.}
\end{figure}

The best fits obtained with each dipole model are compared to the HERA data in
figure \ref{fig:H1-xsec}, \ref{fig:ZEUS-xsec} and \ref{fig:HERA-ratio}. 
The corresponding fitted parameters are given in table
\ref{tab:Best-fit-params}.
Note that we achieve a lower $\chi^2/\mbox{d.o.f}=0.89$ with CGC[$0.74$]
than with CGC[$0.63$] and FSSat for which we obtain $\chi^2/\mbox{d.o.f}=0.96$
and $\chi^2/\mbox{d.o.f}=0.97$ respectively. Compared to
the FSSat and CGC[$0.63$] fits, note that no
additional enhancement in the transverse wavefunction is required in the
CGC[$0.74$] fit. Nevertheless the
extracted wavefunction still exhibits enhancement compared to the old BG
wavefunction. The extracted light-cone wavefunctions are shown in figure
\ref{fig:Psisq_r0}.

\section{Distribution Amplitudes}
\begin{figure}
\begin{center}
\includegraphics[width=0.4\textwidth]{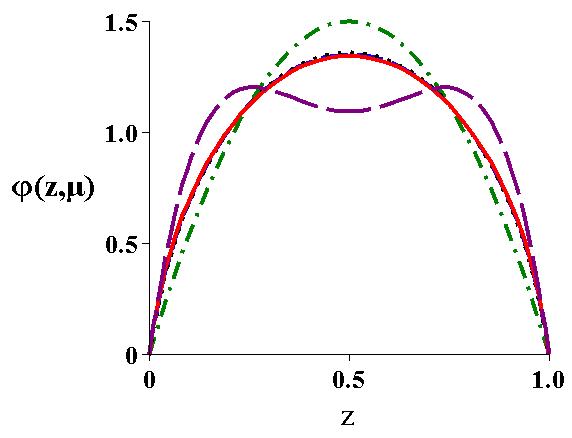}
\end{center}
\caption{\label{fig:DA_1GeV}The extracted leading twist-$2$ DAs at $\mu=1$ GeV
(CGC[$0.74$]: solid;
FSSat: dotted; CGC[$0.63$]: dashed) compared to the DA of reference
~\cite{Ball:1996tb} also at $1$ GeV (long-dashed) and the asymptotic
DA (dot-dashed).}
\end{figure}

The leading twist-$2$ Distribution Amplitude (DA) is given by
~\cite{Forshaw:2010py}
\begin{equation}
\varphi(z,\mu) \sim \left( 1 - \mathrm{e}^{-\mu^2/\Delta(z)^2}\right)
  \mathrm{e}^{-m_f^{2}/\Delta(z)^2} [z(1-z)]^{b_L}~,
\end{equation}
where $\Delta(z)^2 = 8[z(1-z)]^{b_L}/R_L^2$. This leading twist DA is only
sensitive to the longitudinal wavefunction and, as illustrated in figure
\ref{fig:DA_1GeV}, we expect little variation in the predictions using the
different dipole models. 
To compare with existing theoretical predictions for the DA, we
compute moments, i.e.
\begin{equation}
\langle \xi^n \rangle_{\mu} = \int_0^1 \d z \; \xi^n \varphi(z,\mu)~. 
\end{equation}
where by convention~\cite{Forshaw:2010py}
\begin{equation}
\int_0^1 \d z ~\varphi(z,\mu) = 1~.
\end{equation}

In reference~\cite{Forshaw:2010py}, we noted that our DA is very slowly varying
with
$\mu$ for $\mu > 1$~GeV, i.e  our parameterization neglects the
perturbatively known
$\mu$-dependence of the DA. This statement remains true
if we use the CGC[$0.63$] or CGC[$0.74$] instead of the FSSat model. 

Our results are compared with the existing predictions in table
\ref{tab:moments-mu}. The moments
obtained with our best fit, i.e with the CGC[$0.74$] model, are very similar to
those obtained with FSSat model or the CGC[$0.63$]. In all cases, the results
are in very good agreement with expectations based on QCD sum rules and the
lattice.

\begin{table}[h]
\begin{center}
\textbf{Moments of the leading twist DA at the scale $\mu$}
\[
\begin{array} 
[c]{|c|c|c|c|c|c|c|c|c}\hline
\mbox{Reference} & \mbox{Approach} & \mbox{Scale}~\mu &\langle \xi^2
\rangle_{\mu}&\langle \xi^4 \rangle_{\mu}&\langle \xi^6 \rangle_{\mu}&\langle
\xi^8 \rangle_{\mu}&\langle \xi^{10} \rangle_{\mu} \\ \hline
\mbox{(This paper)} & \mbox{CGC[$0.74$] fit}&\sim 1~\mbox{GeV} &0.227 &
0.105& 0.062 &0.041 &0.029\\ \hline
\mbox{(This paper)} & \mbox{CGC[$0.63$] fit}&\sim 1~\mbox{GeV} &0.229 &
0.107& 0.063 &0.042 &0.030 \\ \hline
\mbox{\cite{Forshaw:2010py}} & \mbox{FSSat fit}&\sim 1~\mbox{GeV} &0.227&
0.105&0.062&0.041&0.029\\ \hline
\mbox{(This paper)} & \mbox{Old BG prediction}&\sim 1~\mbox{GeV} &0.181&
0.071&0.036&0.021&0.014\\ \hline
\mbox{\cite{Bakulev:1998pf}}&
\mbox{GenSR}&1~\mbox{GeV}&0.227(7)&0.095(5)&0.051(4)&0.030(2)&0.020(5) \\ \hline
\mbox{\cite{Chernyak:1983ej}}& \mbox{SR}&1~\mbox{GeV} &0.26&0.15 & & &  \\
\hline
\mbox{\cite{Ball:1996tb}} & \mbox{SR}&1~\mbox{GeV} &0.26(4)& & & &   \\ \hline
\mbox{\cite{Ball:2007zt}}&\mbox{SR}&1~\mbox{GeV} &0.254& & & &  \\ \hline
\mbox{\cite{Ball:2004ye}}&\mbox{SR}&1~\mbox{GeV} &0.23\pm^{0.03}_{0.02}&
0.11\pm_{0.02}^{0.03}& & &  \\ \hline
\mbox{\cite{Boyle:2008nj}}&\mbox{Lattice} &2~\mbox{GeV} &0.24(4)& & & &  \\
\hline 
 & 6z(1-z)
&\infty&0.2&0.086&0.048&0.030&0.021 \\ \hline
\end{array}
\]
\end{center}
\caption {Our extracted values for  $\langle \xi^n \rangle_{\mu}$, compared to
predictions based on the QCD sum rules (SR),
Generalised QCD Sum Rules (GenSR) or lattice QCD.}
\label{tab:moments-mu}
\end{table}

Finally, in figure \ref{fig:DA_1GeV} we compare our DAs with that
predicted by Ball and Braun~\cite{Ball:1996tb}, at a scale $\mu=1$ GeV. 
The agreement is reasonable given that in reference~\cite{Ball:1996tb}, the
expansion
in Gegenbauer polynomials is truncated at low order, which is presumably
responsible for the local minimum at $z=1/2$. Certainly all $4$ distributions
distributions are broader than the asymptotic prediction $\sim 6z(1-z)$.

\section{Conclusions}
We have used the current HERA data on diffractive $\rho$ production to extract
information on the $\rho$ light-cone wavefunction. We find that the
corresponding leading twist-$2$ DA is broader than the asymptotic shape
and agrees very well with the expectations of QCD sum
rules and the lattice. We also find that the data
prefer a transverse wavefunction with end-point enhancement although the degree
of such an enhancement is model-dependent.

\section*{Acknowlegdements}

R.S thanks the Institute for Nuclear Theory (INT) and the Facult\'e des \'Etudes
Superieures et de la Recherche (FESR) of the Universit\'e de Moncton for 
financial support. R.S also thanks the organisers for their invitation and for
making this workshop most enjoyable. We thank H. Kowalski and C.
Marquet for useful discussions. This research is also supported by the
UK's STFC.


\end{document}